\documentclass[epj,twocolumn]{webofc}
\usepackage[varg]{txfonts}   
\def\bb    #1{\hbox{\boldmath${#1}$}}

\woctitle{XLIV International Symposium on Multiparticle Dynamics}
\begin{document}
\title{ Possible
Implication of a Single Nonextensive  $p_T$ Distribution  for Hadron Production
in High-Energy $pp$ Collisions
\thanks{Presented by G.Wilk}}
%
%
\author{
       Cheuk-Yin Wong\inst{1}\fnsep\thanks{\email{wongc@ornl.gov}},
       Grzegorz
        Wilk\inst{2}\fnsep\thanks{\email{wilk@fuw.edu.pl}},
        Leonardo J.\ L.\ Cirto\inst{3}\fnsep\thanks{\email{cirto@cbpf.br}},
        Constantino Tsallis\inst{2,4}\fnsep\thanks{\email{tsallis@cbpf.br}}
}

\institute{
Physics Division, Oak Ridge National Laboratory, Oak
Ridge, Tennessee 37831, USA \and
National Centre for Nuclear
Research, Warsaw 00-681, Poland \and
Centro Brasileiro de
Pesquisas Fisicas \& National Institute of Science and Technology
for Complex Systems,\\ \hspace*{0.02cm}~~Rua Xavier Sigaud 150, 22290-180 Rio de
Janeiro-RJ, Brazil \and Santa Fe Institute, 1399 Hyde Park Road,
Santa Fe, NM 87501, USA
          }

\abstract{Multiparticle production processes in $pp$ collisions at the
  central rapidity region are usually considered to be divided into
  independent "soft" and "hard" components.  The first is described by
  exponential (thermal-like) transverse momentum spectra in the
  low-$p_T$ region with a scale parameter $T$ associated with the
  temperature of the hadronizing system. The second is governed by a
  power-like distributions of transverse momenta with power index $n$
  at high-$p_T$ associated with the hard scattering between partons.
  We show that the hard-scattering integral can be approximated as a
  nonextensive distribution of a quasi-power-law containing a scale
  parameter $T$ and a power index $n=1/(q -1)$, where $q$ is the
  nonextensivity parameter.  We demonstrate that the whole region of
  transverse momenta presently measurable at LHC experiments at
  central rapidity (in which the observed cross sections varies by
  $14$ orders of magnitude down to the low $p_T$ region) can be
  adequately described by a single nonextensive distribution.  These
  results suggest the dominance of the hard-scattering
  hadron-production process and the approximate validity of a
  ``no-hair" statistical-mechanical description of the $p_T$ spectra
  for the whole $p_T$ region at central rapidity for $pp$ collisions
  at high-energies.  }
\maketitle
\section{Introduction}
\label{Introduction}

Particle production in $pp$ collisions comprises of many different
mechanisms in different parts of the phase space.  We shall be
interested in particle production in the central rapidity region where
it is customary to divide the multiparticle production into
independent soft and hard processes populating different parts of the
transverse momentum space separated by a momentum scale $p_0$. As a
rule of thumb, the spectra of the soft processes in the low-$p_T$
region are (almost) exponential, $F (p_T )$$\sim$$\exp
(-p_T/T)$, and are usually associated with the thermodynamical
description of the hadronizing system, the fragmentation of a flux
tube with a transverse dimension, or the production of particles by
the Schwinger mechanism \cite{And83,Gat92,Sch51,Wan88,WongHard}. The
$p_T$ spectra of the hard process in the high-$p_T$ region are
regarded as essentially power-like, $F (p_T )$$\sim$$p_T^{-n}$, and
are usually associated with the hard scattering process
\cite{Bla74,TR,Sj,Wan91,Won98}. However, it was found already long
time ago that both description could be replaced by simple
interpolating formula \cite{Michael},
\begin{eqnarray}
F(p_T)=A \left ( 1 + \frac{p_T}{p_0} \right ) ^ {-n}, \label{CM-H}
\end{eqnarray}
that becomes power-like for high $p_T$ and exponential-like for
low $p_T$. Notice that for high $p_T$, where we are usually
neglecting the constant term, the scale parameter $p_0$ becomes
irrelevant, whereas for low $p_T$ it becomes, together with power
index $n$, an effective temperature $T=p_0/n$. The same formula
re-emerged later to become known as the {\it QCD-based Hagedorn
formula} \cite{H}. It was used for the first time in the analysis
of UA1 experimental data \cite{UA1} and it became one of the
standard phenomenological formulas for $p_T$ data analysis.

In the mean time it was realized that Eq. (1) is just
another realization of the nonextensive distribution \cite{T}  with
parameters $q$ and $T$, and a normalization constant $A$,
\begin{equation}
F\left(p_{T}\right)=A~\left[1-\left(1-q\right)\frac{p_{T}}{T}\right]^{1/(1-q)},
\label{T}
\end{equation}
that has been widely used in many other branches of physics. For our
purposes, both formulas are equivalent with the identification of $n$
= $1/(q-1)$ and $p_0=nT$, and we shall use them interchangeably. Because
Eq. (\ref{T}) describes nonextensive systems in statistical mechanics,
the parameter $q$ is usually called the {\it nonextensivity
  parameter}. As one can see, Eq. (2) becomes the usual
Boltzmann-Gibbs exponential distribution for $q$ $\to$ 1, with $T$
becoming the temperature.  Both Eqs. (1) and (2) have been widely used
in the phenomenological analysis of multiparticle productions (cf.,
for example
\cite{BCM,Beck,RWW,PHENIX,STAR,CMS,ATLAS,ALICE,Wibig,B_et_all,JCleymans,ADeppman,Others,WalRaf})\footnote{
  For those who would like to use Eq. (2) in the context of
  nonextensive thermodynamics (as is done, for example, in
  \cite{JCleymans,ADeppman}) references in \cite{TherCons} provide
  arguments that this is fully legitimate. Outside the physics of
  multiparticle production, this approach is much better known and
  commonly used (see for example, \cite{Contemporary,T} for details
  and references).}.

We shall demonstrate here that, similar to the original ideas
presented in \cite{Michael,H}, the whole region of transverse
  momenta presently measurable at LHC experiments (which spans now
  enormous range of $\sim$14 orders of magnitude in the measured
  cross-sections down to the low-$p_T$ region) \cite{CMS,ATLAS,ALICE}
  can be adequately described by a {\it single quasi-power law
    distribution}, either Eq. (\ref{CM-H}) or Eq. (\ref{T}).  We shall
  offer a possible explanation of this phenomenon by showing
   that the hard-scattering integral can be cast approximately into a
  non-extensive distribution form and that the description of a single nonextensive $p_T$ distribution  for the $p_T$ spectra over the whole $p_T$ region
suggests the dominance of the hard-scattering process  at central rapidity   for high-energy $pp$ collisions. 

\section{Questions associated with a Single Nonextensive distribution  for $p_T$ spectra in pp collisions} \label{sec:CYWW}

The possibility of two components in the transverse spectra implies
that its complete description will need two independent functions with
different sets of parameters, each dominating over different regions
of the transverse momentum space. The presence of two different
components will be indicated by gross deviations when the spectrum
over the whole transverse space is analyzed with only a single
component.  An example for the presence of two (or more) components of
production processes can be clearly seen in Fig. 1 of \cite{ALI13}, in
the $p_T$ spectra in central (0-6\%) PbPb collisions at $\sqrt{s_{NN}}
$=2.76 TeV from the ALICE Collaboration, where two independent
functions are needed to describe the whole spectra as described in
\cite{Urm14,MR}.

 For our purposes in studying produced hadrons in $pp$ collisions,
 where the high $p_T$ hard-scattering component is expected to have a
 power-law form with a power index $n$, either (1) or (2) can be
 written as
\begin{equation}
E\frac{d\sigma}{d^3p}=\frac{A}{\left ( 1+\frac{m_T - m}{nT}
\right )^n},
\end{equation}
where $n$ is the power index, $T$ is the `temperature' parameter, and
$m$ and $m_T$=$\sqrt{m^2+p_T^2}$ are the rest mass and transverse mass
of the produced hadrons which are taken to be the dominant particles,
the pions. It came as a surprise to us that for $pp$ collisions at
$\sqrt{s_{NN}}$ = 7 TeV, the $p_T$ spectra within a very broad range,
from 0.5 GeV up to 181 GeV, in which cross section varies by 14 orders
of magnitude, can still be described well by a single nonextensive
formula with power index $n$ = 6.6 \cite{Won12}.  The good fits to the
$p_T$ spectra over such a large range of $p_T$ with only three
parameters, $(A, n, T)$, raise intriguing questions :

\begin{itemize}
\item Why are there only three degrees of freedom in the spectra over
  such a large $p_T$ domain?  Does it imply that there is only a
  single component, the hard scattering process, contributing
  dominantly over the whole $p_T$ domain?  If so, are there supporting
  experimental evidences from other correlation measurements?

\item Mathematically, the power index $n$ is related to the parameter
  $q = 1 + 1/n$ in non-extensive statistical mechanics \cite{T}.  What
  is the physical meaning of $n$? If $n$ is related to the power index
  of the parton-parton scattering law, then why is the observed value
  so large, $n$$ \sim$7, rather than $n \sim 4$ as predicted naively by
  pQCD?

\item Are the power indices for jet production different from
those for hadron production?  If so, why ?

\item Do multiple parton collisions play any role in modifying the
power index $n$?

\item In addition to the power law $1/p_{{}_T}^n$, does the
  differential cross section contain other additional
  $p_{{}_T}$-dependent factors?  If they are present, how do they
  change the power index?

\end{itemize}

These questions were discussed and, at least partially, answered in
\cite{Won13}.  Before proceeding to our main point of phenomenological
considerations we shall first recapitulate briefly the main results of
this attempt to reconcile, as far as possible, the nonextensive
distribution with the QCD where, as shown in \cite{Won13}, the only
relevant ingredients from QCD are hard scatterings between
constituents resulting in the production of jets which further undergo
fragmentation, showering, and hadronization to become the observed
hadrons.

\section{Approximate Hard-Scattering Integral \label{sec2}}

The answers to the questions posed above will be facilitated with an
approximate analytical form of the hard-scattering integral. We start
with the relativistic hard-scattering model as proposed in
\cite{Bla74}\footnote{For the history of the power law, see
  \cite{TR}.}  and examined in \cite{WongHard,Won98,Won13}. We
consider the collision of projectiles $A$ and $B$ in the
center-of-mass frame at an energy $\sqrt{s}$ in the reaction $A + B
\to c + X$, with $c$ coming out at midrapidity, $\eta$$\sim$ 0. Upon
neglecting the intrinsic transverse momentum and rest masses, the
differential cross section in the lowest-order parton-parton elastic
collisions is given by
\begin{eqnarray}
\hspace*{-0.8cm}\frac{E_cd^3\sigma( AB \to c X) }{dc^3}\!\!
&=&\!\!\!\!\! \sum_{ab} \int dx_a
dx_b G_{a/A}(x_a) G_{b/B} (x_b)\nonumber\\
&& \times
 \frac{E_cd^3\sigma( ab \to c X') }{dc^3}.
\end{eqnarray}
The parton-parton invariant cross section is related to $d\sigma(
ab\!\! \to\!\! c X') /dt$ by
\begin{eqnarray}
\hspace{ -5mm} E_c \frac{d^3\sigma( ab \to c X') }{dc^3} =
 \frac{\hat s}{\pi}\frac{
  d\sigma( ab \to cX') } {dt} \delta (\hat s +\hat t +\hat u ),
\label{6}
\end{eqnarray}
where
\begin{eqnarray}
\hat s = (a+b)^2,~~ \hat t = (a-b)^2,~~ \hat u =
(b-c)^2.\label{6a}
\end{eqnarray}
In the infinite momentum frame the momenta can be written as
\begin{eqnarray}
a& = & \left(x_a \frac{\sqrt{s}}{2}, ~{\bb O}_T, ~x_a
\frac{\sqrt{s}}{2} \right),\nonumber\\
b& = & \left(x_b \frac{\sqrt{s}}{2} , ~{\bb O}_T, -x_b
\frac{\sqrt{s}}{2} \right),
\nonumber\\
c& = & \left( x_c \frac{\sqrt{s}}{2} + \frac{c_T^2}{2x_c
\sqrt{s}}, ~{\bb c}_T, ~x_c \frac{\sqrt{s}}{2} - \frac{c_T^2}{2x_c
\sqrt{s}}\right) .\nonumber
\end{eqnarray}
We denote light-cone variable $x_c$ of the produced parton c as
$x_c = { \left( c_0+c_z \right)}/{\sqrt{s}}$. The constraint of
$\hat s +\hat t + \hat u=0$ gives
\begin{eqnarray}
&&x_a \left( x_b \right) = x_c + \frac{c_T^2}{ \left( x_b
 -\frac{  c_T^2}{x_c s}\right)s}.
\end{eqnarray}
We consider only the special case of $c$ coming out at $\theta_c=
90^o$, in which $x_c = \frac{c_T}{\sqrt{s}}$, $x_a(x_b) = x_c +
x_c^2/\left(x_b - x_c\right)$ and $x_a = x_b = 2x_c$. We have
therefore
\begin{eqnarray}
\hspace*{-0.8cm}\frac{E_c d^3\sigma( \!AB\! \!\to \! c X\!) }{dc^3}\biggr |_{y\sim 0}
\!\!&=&\!\!\sum_{ab}\!\!\! \int \!\!\! dx_b dx_a G_{a/A}(x_a\!)
G_{b/B} (x_b\!) \nonumber\\
& \times&\!\!\! \frac{x_a x_b \delta (x_a\! -\! x_a(x_b))}{\pi
(x_b-c_T^2/x_c s)} \frac{d\sigma( ab\! \to\! cX'\!)}{dt},\nonumber
\end{eqnarray}
where ${\cal G}_a(x_a) = x_a G_{a/A}(x_a)$ and ${\cal
G}_b(x_b)=x_a G_{b/B}(x_b)$.  After integrating over $x_a$, we
obtain
\begin{eqnarray}
\hspace*{-0.8cm}\frac{E_Cd^3\sigma( AB \to c X) }{dc^3}\biggr |_{y\sim 0}
&=&\sum_{ab}\int dx_b \frac{{\cal G}_a(x_a(x_b)) {\cal
G}_b(x_b)}{\pi (x_b-c_T^2/x_c s)}\nonumber\\
&\times&\!\!\! \frac{d\sigma(ab\! \to\! cX')}{dt}.
\end{eqnarray}
To integrate over $x_b$, we use the saddle point method, write
${\cal G}_a(x_a(x_b)) {\cal G}_b(x_b) = e^{f(x_b)}$, and expand
$f(x_b)$ about its minimum at $x_{b0}$.  We obtain then that
\begin{eqnarray}
\hspace*{-1.cm}\int d x_b e^{f(x_b)} g(x_b) \sim e^{f(x_{b0})} g(x_{b0})
\sqrt{ \frac{2 \pi}{-\partial ^2 f (x_{b}) /\partial
x_b^2|_{x_b=x_{b0}}} }.
\end{eqnarray}
For simplicity, we assume $G_{a/A}$ and $G_{b/B}$ to have the same
form.  At $\theta_c\sim 90^0$ in the CM system, the minimum value
of $f(x_b)$ is located at 
\begin{eqnarray}
 x_{b0}=x_{a0}=2x_c,
\label{eq10}
\end{eqnarray}
and we get the hard-scattering integral 
\begin{eqnarray}
\hspace*{-0.8cm}E_C \frac{d^3\sigma( AB \to c X) }{dc^3}\biggr |_{y\sim 0}\!\!\!\!\!\!\!
&\sim&\!\!\!\!\! \sum_{ab} B [x_{a0}G_{a/A}(x_{a0})][
x_{b0}G_{b/B}(x_{b0})] \nonumber \\
&\times& \frac{d\sigma(ab\! \to\! cX')}{dt}
\end{eqnarray}
where
\begin{eqnarray}
\!\!\!\!\!B &=&\frac{1}{\pi (x_b-c_T^2/x_c s)} \sqrt{ \frac{2
\pi}{-\partial ^2 f (x_{b}) /\partial x_b^2|_{x_b=x_{b0}}} }.
\end{eqnarray}
 For the case of ${\cal
G}_a(x_a)=x_aG_{a/A}(x_a)=A_a(1-x_a)^{g_a}$, we find
\begin{eqnarray}
\hspace*{-1.cm} E_C \frac{d^3\sigma( AB\!\! \to\!\! c X)
}{dc^3}\biggr |_{y\sim 0} \!\!\!\!\!&\sim& \!\!\! \!\!\sum_{ab}{ A_aA_b}
\frac{(1-x_{a0})^{g_a+\frac{1}{2}}(1-x_{b0})^{g_b+\frac{1}{2}}}
{\sqrt{\pi g_a}\sqrt{ x_c(1-x_c)}}\nonumber\\
 &\times& \frac{d\sigma(ab\! \to\! cX')}{dt}.
\end{eqnarray}
If the basic process $ab \to cX'$ is $gg \to gg$ or $ab \to cX'$
is $qq' \to qq'$, the cross sections at $\theta_c\sim 90^{o}$
\cite{Gas90} are
\begin{eqnarray}
\frac{d\sigma(gg\to gg) }{dt} &\sim & \frac{9\pi
\alpha_s^2}{16c_T^4} \left [\frac{3}{2} \right
]^3,\nonumber\\
\frac{d\sigma(qq' \to qq')}{dt}& \sim & \frac{4 \pi
\alpha_s^2}{9c_T^4} \frac{5}{16}. \label{38}
\end{eqnarray}
In both cases, the differential cross section behave as
$d\sigma(ab\!\!\to\!\! cX')/dt \sim \alpha_s^2/(c_T^2)^2$.

\section{Parton Multiple  Scattering \label{sec3}}

As the collision energy increases, the value of $x_c$ gets smaller
and the number of partons and their density increase rapidly. Thus
the total hard-scattering cross section increases as well
\cite{Sj}. The presence of a large number of partons in the
colliding system results in multiple hard-scatterings of
projectile parton on partons from target nucleon.  

We find that for the process of $a\to c$ in the collision of a
parton $a$ with a target of $A$ partons in sequence without a
centrality selection, the $\bb c_{{}_T}$-distribution is given by
\cite{Won13}
\begin{eqnarray}
&&\hspace*{-1.3cm}\frac{d \sigma_{H}^{(tot)} (a\to c)}{d \bb c_T} ~= ~ A
\frac{\alpha_s^2}{ c_T^4} \int d\bb b ~T(b)
\\
&&\hspace*{-0.8cm}+ \frac{A(A-1)}{2} \frac{16\pi\alpha_s^4}{c_T^6}\cdot
 \ln \{\frac{c_T}{2p_0}\} \int d\bb b[T(b)]^2\cdot
\nonumber\\
&& \hspace*{-0.8cm}+\frac{A(A-1)(A-2)}{6} \frac{936\pi^2\alpha_s^6
}{c_T^8} [ \ln \frac{c_T}{3p_0}]^2 \int d\bb b[T(b)]^3, \nonumber
\label{eq46}
\end{eqnarray}
where the terms on the right-hand side correspond to collisions of
the incident parton with one, two and three target partons,
respectively.   Here, the quantity $A$ is the number of partons in the 
nucleon as a composite system and is the  integral of the parton
density over the parton momentum fraction. This result shows that
without centrality selection in minimum-biased events, the
differential cross section will be dominated by the contribution
from a single parton-parton scattering that behaves as
$\alpha_s^2/c_T^4$ (cf. previous analysis on the multiple
had-scattering process in \cite{Kas87,Cal90,Gyu01}). Multiple
scatterings with $N>1$ scatterers contribute to terms of order
$\alpha_s^{2N}$ $[\ln{(C_T/Np_0)}]^{N-1}/c_T^{2+2N}$ \cite{Won13}.

\section{The Power Index in Jet Production \label{sec4}}

From the above results one gets the approximate analytical formula
for hard-scattering invariant cross section $\sigma_{\rm  inv}$,
for $A+B \to c+X$ at midrapidity, $\eta\sim 0$, equal to
\begin{eqnarray}
  \hspace*{-1.0cm}E_c \frac{d^3\sigma( AB\!\! \to\!\! c X)
}{dc^3}
\biggr |_{y\sim 0}
\!\!\!\!\!\!\!\propto\!\!
 \frac{\alpha_s^2 (1\!-\!x_{a0}(c_T))^{g_a+\frac{1}{2}}(1\!-\!x_{b0}(c_T))^{g_b+\frac{1}{2}}}
{c_T^{4}\sqrt{c_T/\sqrt{s}}  \sqrt{1-x_c}}\!. \label{19}
\end{eqnarray}

The power index $n$ has here the value $4 + 1/2$.  Its value can
be extracted by plotting  $(\ln{ \sigma_{\rm inv}}) $ as a
function of $(\ln c_T)$ (then the slope in the linear section gives
the value of $n$, and the variation of $(\ln{ \sigma_{\rm inv}} )$
at large $(\ln c_T)$ gives the value of $g_a$ and $g_b$).   One can also 
consider for this purpose a fixed $x_c$ and look at two different
energies (as suggested in \cite{Arl10}),
\begin{eqnarray}
\frac{\ln[\sigma_{\rm inv}(\sqrt{s_1},x_c)/\sigma_{\rm
inv}(\sqrt{s_2},x_c)]}{\ln[\sqrt{s_2}/\sqrt{s_1}]} \sim
n(x_c)-\frac{1}{2}.
\end{eqnarray}
We follow an alternative method and analyze the $p_{{}_T}$ spectra
using a running coupling constant,
\begin{eqnarray}
\alpha_s(Q^2(c_T)) = \frac{12\pi}{27 \ln(C + Q^2/\Lambda_{\rm
QCD}^2)}, \label{run}
\end{eqnarray}
where we have chosen $\Lambda_{\rm QCD}$ to be 0.25 GeV to give
$\alpha_s(M_Z^2)=0.1184$ \cite{Ber12}.  We identify $Q$ as $c_{{}_T}$
and have chosen $C$=10 both to give $\alpha_s(Q$$\sim$$\Lambda_{\rm
  QCD})$ $\sim$ 0.6 in hadron spectroscopy studies \cite{Won01} and to
regularize the coupling constant for small values of $Q(c_{{}_T})$.
We search for $n$ by writing the invariant cross section
Eq.\ (\ref{19}) for jet production as
\begin{eqnarray}
&& \hspace*{-1.6cm}  E_c \frac{d^3\sigma( AB\!\! \to\!\! c X) }{dc^3} \biggr |_{y\sim 0}\nonumber\\
&&  \hspace*{-1.0cm}\propto
 \frac{\alpha_s^2(Q^2(c_T)) (1-x_{a0}(c_T))^{g_a+\frac{1}{2}}(1-x_{b0}(c_T))^{g_b+\frac{1}{2}}}
{c_T^{n} \sqrt{1-x_c}}. \label{22}
\end{eqnarray}

\begin{figure}[t]
\hspace*{-0.4cm}
\includegraphics[scale=0.33]{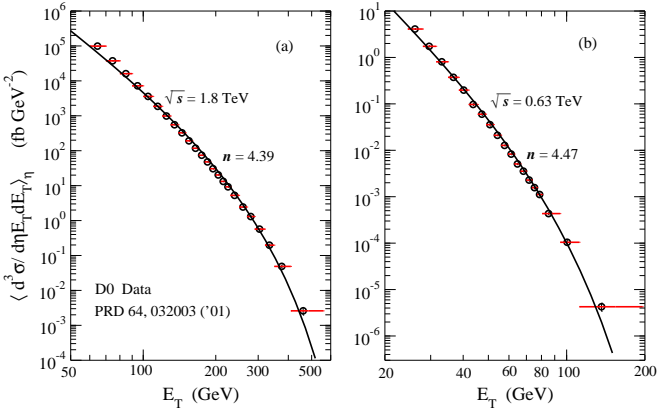}
\caption{(Color online) Comparison of the relativistic hard-scattering
  model results for jet production, Eq.\ (\ref{22}) (solid curves),
  with experimental $d\sigma/d\eta E_T dE_T$ data from the D0
  Collaboration \cite{Abb01}, for hadron jet production within
  $|\eta|$$<$0.5, in $\bar p p$ collision at (a) $\sqrt{s}$=1.80 TeV,
  and (b) $\sqrt{s}$=0.63 TeV. } \label{1}
\end{figure}
\vspace*{-0.4cm}

\begin{table}[h]
\hspace*{-2.5cm}\caption {The power index for jet production in $\bar p
p$ and $pp$  collisions } 
\begin{tabular}{|c|c|c|c|c|}
\cline{1-5}
 $\!\!\!$Collaboration$\!\!\!$&  $\sqrt{s}$ & $R$  &  $\eta$   & $n$
  \\
\cline{1-5}
     D0 \cite{ Abb01}          & $\bar p p$ at 1.80 TeV&  0.7  &     $|\eta|<$  0.7  & 4.39
 \\ \hline
     D0 \cite{ Abb01}            & $\bar p p$ at 0.63 TeV&  0.7  &  $|\eta|<$  0.7   & 4.47
 \\ \hline
     ALICE   \cite{Alice13}    & $p p$ at   2.76 TeV&  0.2  & $|\eta|<$  0.5  & 4.78
 \\ \hline
     ALICE  \cite{Alice13}     & $p p$ at 2.76 TeV&  0.4  &  $|\eta|<$  0.5  & 4.98
  \\ \hline
     CMS \cite{cms11}        & $p p$ at 7 TeV&  0.5  &  $|\eta|<$ 0.5  & 5.39
 \\ \hline
\end{tabular}
\end{table}

\vspace*{0.3cm}
\noindent
In the literature \cite{Duk84,ZEUS} the index $g_a$ for the
structure function of a gluon varies from 6 to 10. Following
\cite{Duk84} we shall take $g_a=6$.  As shown in Fig.\ 1 and Table I,
data from D0 \cite{Abb01} on $d\sigma/d\eta E_T dE_T$ for hadron
jet production within $|\eta|$$<$0.5 can be fitted with $n$=4.39
for $\bar p p$ collisions at $\sqrt{s}$=1.8 TeV, and with $n$=4.47
for $\bar p p$ collisions at $\sqrt{s}$=0.630 TeV.  In other
comparisons with the ALICE data for jet production in $pp$
collisions at $\sqrt{s}=2.76$ TeV at the LHC within $|\eta|<0.5$
\cite{Alice13}, the power index is $n$=4.78 for $R=0.2$, and is
$n$=4.98 for $R=0.4$ (Table I).  The power index is $n$=5.39, for
CMS jet differential cross section in $pp$ collisions at
$\sqrt{s}=7$ TeV at the LHC within $|\eta|<0.5$ and $R=0.5$
\cite{cms11}.  This latter $n$ value exceeds slightly the expected
value of $n=4.5$.

Except for the CMS data at 7 TeV that may need further
re-examination, the power indices extracted for hadron jet
production and listed in Table I are in approximate agreement with
the value of $n$=4.5 in Eq.\ (\ref{19}) and with previous analysis
of Arleo $et~al.$ \cite{Arl10}, indicating the approximate
validity of the hard-scattering model for jet production in
hadron-hadron collisions, with the predominant $\alpha_s^2/c_T^4$
parton-parton differential cross section as predicted by pQCD.

\section{Change of the Power Index $n$ from Jet Production to   Hadron Production}

The results in the last section indicates that the simple
hard-scattering model, i.e., Eq. (18), adequately describes the power
index of $n \sim$ 4.5 for jet production in high-energy $pp$
collisions.  However, the power index for hadron production is
considerable greater, in the range of $n$$\sim 6-10$
\cite{Won12,Arl10}.  What is the origin of the increase in  the power
index $n$?

A jet $c$ evolves by fragmentation, showering, and hadronization to
turn the jet into a large numbers of hadrons in a cone along the jet
axis.  The showering of the partons will go through many generations
of branching.  If we label the (average) momentum of the $i$-th
generation parton by $p_T^{(i)}$, the showering can be represented as
$c_T \to p_T^{(1)} \to p_T^{(2)} \to p_T^{(3)} \to ...\to
p_T^{(\lambda)}$.  Each branching will kinematically degrade the
momentum of the showering parton by a momentum fraction, $\zeta$=$
p_T^{(i+1)}/ p_T^{(i)}$.  At the end of the terminating $\lambda$-th
generation of the showering, and hadronization, the $p_T$ of a
produced hadron is related to the $c_T$ of the parent parton jet by
\begin{eqnarray}
\frac{p_T}{c_T} \equiv \frac{p_T^{(\lambda)}}{c_T} =\zeta^\lambda  .
\label{eq52}
\end{eqnarray}
It is easy to prove that if the generation number
$\lambda$ and the fragmentation fraction $z$ are  independent of
the jet $c_T$, then the power law and the power index  for the
$p_T$ distribution are unchanged \cite{Won13}.

 We note however that in addition to the kinematic decrease of $p_T$
 as described by (\ref{eq52}), the showering generation number
 $\lambda$ is governed by an additional criterion on the virtuality,
 which measures the degree of the off-the-mass-shell property of the
 parton.  From the different parton showering schemes in the PYTHIA
 \cite{PYTHIA}, the HERWIG \cite{HERWIG}, and the ARIADNE
 \cite{ARIADNE}, we can extract a general picture that the initial
 parton with a large initial virtuality $Q$ decreases its virtuality
 by showering until a limit of $Q_0$ is reached.  The downgrading of
 the virtuality will proceed as $Q$=$Q^{(0)} \to Q^{(1)} \to Q^{(2)}
 \to Q^{(3)} \to ... \to Q^{(\lambda)}$=$Q_0$.  There is a one-to-one
 mapping of the initial virtuality $Q$ with the transverse momentum
 $c_T$ of the evolving parton as $Q(c_T)$ (or conversely $c_T(Q)$).
 Because of such a mapping, the decrease in virtuality $Q$ corresponds
 to a decrease of the corresponding mapped $\tilde c_T$ as
 $c_T$=$\tilde c_T^{(0)} \to \tilde c_T^{(1)} \to \tilde c_T^{(2)} \to
 \tilde c_T^{(3)} \to ...\to \tilde c_T^{(\lambda)}$=$ c_T(Q_0)$,
 where $\tilde c_T^{(i)}$=$c_T(Q^{(i)})$.  The cut-off virtuality
 $Q_0$ maps into a transverse momentum $c_{T0}$=$c_T(Q_0)$.  In each
 successive generation of the showering, the virtuality decreases by a
 virtuality fraction which corresponds, at least approximately, in
 terms of the corresponding mapped parton transverse momentum $\tilde
 c_T^{(i)}$, to a decrease by a corresponding transverse momentum
 fraction, $\tilde \zeta$=$\tilde c_T^{(i+1)}/\tilde c_T^{(i)}$.  The
 showering will end in $\lambda$ generations such that
\begin{eqnarray}
\frac{ c_{T0}}{c_T} \equiv  \frac{\tilde c_{T}(Q^{(\lambda)})} {c_T}=\tilde \zeta^\lambda,
\label{20}
\end{eqnarray}
We can infer a relation
between $c_T$ and the number of generations, $\lambda$,
\begin{eqnarray}
\lambda=\ln \left ( \frac{c_{T0}} {c_T}  \right ) \biggr / {\ln \tilde \zeta}.
\end{eqnarray}
Thus, the showering generation number
$\lambda$ depends on the magnitude of $c_T$.  
On the other hand, kinematically, the showering processes degrades the
transverse momentum of the parton $c_T$ to that of the $p_T$ of the
produced hadron as given by Eq.\ (\ref{eq52}), depending on the number
of generations $\lambda$.  The magnitude of the transverse momentum
$p_T$ of the produced hadron is related  to the
transverse momentum $c_T$ of the parent parton jet by
\begin{eqnarray}
\frac{p_T}{c_T} = \zeta^{\lambda}= \zeta ^{{\ln \frac{c_{T0}}{c_T} } /{\ln \tilde \zeta}}.
\end{eqnarray}
We can solve the above equation for $p_T$ as a function of $c_T$ and obtain
\begin{eqnarray}
\hspace*{-0.5cm}\frac{p_T}{c_{T0}}
=\left ( \frac{c_T}{c_{T0}} \right )
 ^{1- \mu},  ~~{\rm and~~}
\frac{c_T}{c_{T0}}
=\left ( \frac{p_T}{c_{T0}} \right )
 ^{1/(1- \mu)},
\label{23}
\end{eqnarray}
where
\begin{eqnarray}
\mu=\ln \zeta /{\ln \tilde \zeta},
\label{24a}
\end{eqnarray}
and $\mu$ is a parameter that can be searched to fit the data.  As a
result of the virtuality ordering and virtuality cut-off, the hadron
fragment transverse momentum $p_T$ is related to the parton momentum
$c_T$ nonlinearly by an exponent $1-\mu$.

After the fragmentation and showering of the parent parton $c_T$ to the
produced hadron $p_T$, the hard-scattering cross section for the
scattering in terms of hadron momentum $p_T$ becomes
\begin{eqnarray}
&&\hspace*{-1.5cm}
 \frac{d^3\sigma( AB \to p X) }{dy d{\bb p}_T}
= \frac{d^3\sigma( AB \to c X) }{dy d{\bb c}_T}
\frac{d{\bb c}_T}{d{\bb p}_T}
\end{eqnarray}
Upon substituting the non-linear relation (\ref{23}) between the
parent parton moment $c_T$ and the produced hadron $p_T$ in
Eq.\ (\ref{23}), we get
\begin{eqnarray}
\frac{d{\bb c}_T}{d{\bb p}_T}
=
{\frac{1}{1- \mu}}\left ( \frac{p_T}{c_{T0}} \right )
 ^{\frac{2\mu}{1- \mu}}.
\label{77}
\end{eqnarray}
Therefore under the fragmentation from $c$ to $p$, the hard-scattering
cross section for $AB \to p X$ becomes
\begin{eqnarray}
&& \hspace*{-1.4cm}  E_c \frac{d^3\sigma( AB\!\! \to\!\! p X) }{dp^3}\biggr |_{y\sim 0}
=\frac{d^3\sigma( AB \to p X) }{dy d{\bb p}_T}\biggr |_{y\sim 0}  \nonumber\\
&&  \!\!\!\!\!\!\!\!\!\!\!\!\!\!\!\!\!\!\!\! \propto
 \frac{\alpha_s^2(Q^2(c_T)) (1-x_{a0}(c_T))^{g_a+\frac{1}{2}}(1-x_{b0}(c_T))^{g_b+\frac{1}{2}}}
{p_T^{n'} \sqrt{1-x_c(c_T)}}, \label{27}
\end{eqnarray}
where
\begin{eqnarray}
n'=\frac{n-2\mu}{1-\mu},~~{\rm with~~} n=4+\frac{1}{2}.
\end{eqnarray}
Thus, the power index $n$ for jet production can be significantly
changed to $n'$ for hadron production because the greater the value of
the parent jet $c_T$, the greater the number of generations $\lambda$
to reach the produced hadron, and the greater is the kinematic energy
degradation.  By a proper tuning of $\mu$, the power index can be
brought to agree with the observed power index in hadron production.
For example, for $\mu$=0.4 one gets $n'$=6.2 and for $\mu=0.6$ one
gets $n'$=8.2.  Because the parton branching probability, parton
kinematic degradation, and parton virtuality degradation depend on the
coupling constant and the coupling constant depends on the parton
energy, we expect the quantity $\mu$ to depend on the $pp$ collision
energy.  Consequently, $n'$ may change significantly with the
collision energy.

\section{Regularization of the Hard-Scattering Integral}

The power-law (\ref{27}) has been obtained for high $p_T$.  In order
to apply it to the whole range of $E_T$, we need to regularize it
by the replacement,
\begin{equation}
\frac{1}{p_T}  \to \frac{1}{ 1+{m_T}/{m_{T0}}}.
\label{replacement1}
\end{equation}
 The quantity
$m_{T0}$ measures the average transverse mass of the detected hadron
in the hard-scattering process.  The differential cross section
$d^3\sigma( AB \to p X) /dy d{\bb p}_T$ in (\ref{27}) is then
regularized as
\begin{eqnarray}
\!\!\!\!\!\!\!\!\!\!\!\!\!\!\!  \frac{d^3\sigma( AB \to p X) }{dy
d{\bb
p}_T} \biggr | _{ y \sim 0} && \nonumber\\
&& \hspace{-3.0cm} \propto \frac{\alpha_s^2(Q^2(
c_T))(1\!-\!x_{a0}( c_T))^{g_a+1/2}(1\!-\!x_{b0}(
c_T))^{g_b+1/2} } {[1+m_{T}/m_{T0}]^{n'}\sqrt{1-x_c( c_T)}}
. \label{30}
\end{eqnarray}
In the above equation for the production of a hadron with a transverse
momentum $p_T$, the variable $c_T(p_T) $ refers to the transverse
momentum of the parent jet $c_T$ before fragmentation.  We can relate
$p_T$ with $c_T$ by using the empirical fragmentation function of
Ref.\ \cite{BKK} and we get \cite{Won13}
\begin{eqnarray}
c_T(p_T) \sim  p_T \langle \frac{1}{ z} \rangle=  2.33  ~p_T.
\label{eq65}
\end{eqnarray}
This can be regarded as a linearized approximation of Eq. (\ref{23}), which
       shows that $c_T$ and $p_T$ are non-linearly related, when
        we consider the virtuality in the fragmentation process.
\begin{figure}[h]
\centering
 \includegraphics[scale=0.43]{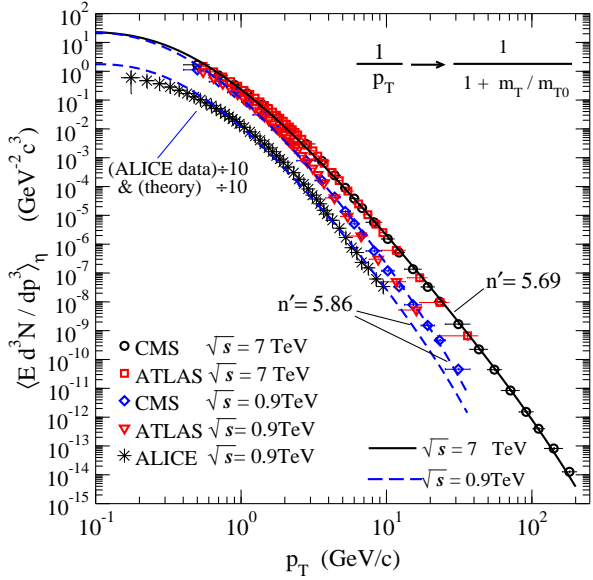}
\vspace*{-0.3cm} \caption{(Color online) Comparison of the
  experimental $\langle E_p d^3N/dp^3\rangle_{\eta}$ data for hadron
  production in $pp$ collisions at the LHC with the relativistic
  hard-scattering model results of (solid and dashed curves)
  Eq.\ (\ref{30}).  The solid line is for $\sqrt{s}$=7 TeV, and the
  dashed line is for $\sqrt{s}$=0.9 TeV.}
 \label{F3}
\end{figure}
Comparisons of the theoretical results calculated with
Eq.\ ({\ref{30}) with the experimental hadron transverse momentum
  distributions in $pp$ collisions at the LHC from the CMS \cite{CMS},
  ATLAS \cite{ATLAS}, and ALICE Collaborations \cite{ALICE} are shown
  in Fig. \ref{F3}.  We find that the experimental data gives
  $n'$=5.69 and $m_{T0}$=0.804 GeV for $\sqrt{s}$=7 TeV and $n'$=5.86
  and $m_{T0}$=0.634 GeV for $\sqrt{s}$=0.9 TeV.  
This indicates that there is indeed a systematic change of the power index $n$
  from jet production to a larger value $n'$ in  hadron production. 
The fits to the low $p_T$ region for the ALICE data can be improved, with
a larger power index $n'$ as we shall see below in Section 9. 

\section{Further Approximation of the Hard-Scattering Integral}
\label{sec:C}

 We would like
to simplify further the $p_T$ dependencies of the structure function
in Eq.\ (\ref{30}) and the running coupling constant as additional
power indices in such a way that will facilitate subsequent
phenomenological comparison.
For parton $c$ coming at mid-rapidity, the quantities $x_{a0}$,
$x_{b0}$, and $x_c$ in  Eqs.\  (\ref{eq10}) and  (\ref{30}) are
\begin{eqnarray} 
x_{a0}=x_{b0}=2x_c, ~~~{\rm and ~~~}
x_c=\frac{c_T}{\sqrt{s} } .
\end{eqnarray}
The structure function factor and the denominator factor in
Eq.\ (\ref{30}) can be approximated for high energies with
$\sqrt{s}\gg c_T$ as
\begin{eqnarray}
\hspace*{-0.9cm}\frac{(1\!-\!x_{a0}( c_T))^{g_a+1/2}(1\!-\!x_{b0}(
c_T))^{g_b+1/2}}{\sqrt{1-x_c(c_T)}}
\sim  {(1-x_c(c_T))^{2 g_a +3/4} } . \nonumber
\end{eqnarray}
 We can relate $c_T$ with $p_T$ by Eq.\ (\ref{eq65})
and 
further approximate the
right-hand side of the above equation in a form that is
advantageous for subsequent purposes.  For high energy with large
$\sqrt{s}$, we make the approximation
\begin{eqnarray}
\hspace*{-0.9cm} {(1\!-\!x_c(c_T))^{2 g_a +\frac{3}{4}} }=\left (1\!-\!\frac{2p_T} {\sqrt{s}}
 \langle \frac{1}{z} \rangle  \right )^{2g_a+3/4}
\hspace*{-0.4cm}\sim \frac{1}{[1+m_T/m_{T0}] ^{n_g}},\!\!
\end{eqnarray}
where 
\begin{eqnarray}
n_g = \frac{2(2g_a+3/4) m_{T0}}  { \sqrt{s}} \langle \frac{1}{z} \rangle   .
\end{eqnarray}
We therefore estimate that $n_g\sim$ 0.04 and 0.007 for $\sqrt{s}=0.9$ and 7 TeV respectively. 

The running coupling constant $\alpha_s$ is a monotonically 
decreasing function of $Q(c_T)$.   It can  be  written approximately as 
\begin{eqnarray}
\alpha_s(Q^2(c_T)) \propto \frac{1}{[1+m_T/m_{T0}] ^{n_\alpha}},
\end{eqnarray}
where $n_\alpha$ can be chosen to minimize errors by matching
$\alpha_s$ at two points of $p_T$.  If we match $\alpha_s(p_T)$ at
$p_T$=$\Lambda_{\rm QCD}$=0.25 GeV and at $p_T$=100 GeV, then
$n_\alpha=$ 0.36.  If we match $\alpha_s$ at $p_T$=$\Lambda_{\rm QCD}$
and at $p_T$=20 GeV, then $n_\alpha=$ 0.46.
 
As a consequence of the above simplifying approximations, we can write
the hard-scattering integral Eq.\ (\ref{30}) in the approximate form
\begin{eqnarray}
\hspace*{-0.9cm} \frac{d^3\sigma( AB \to p X) }{dy
d{\bb p}_T}
\biggr |_{y\sim 0}=F(p_T) 
\sim \frac{A}  {[1+{m_{T}}/{m_{T0}}]^{n}}
, \label{40}
\end{eqnarray}
where
\begin{eqnarray}
n=n'+n_g+n_\alpha,
\end{eqnarray}
and $n'$ is the power index after taking into account the
fragmentation process, $n_g$ the power index from the structure
function, and $n_\alpha$ from the coupling constant.  We note that the
predominant change of the power index from jet production to hadron
production arises from the fragmentation process because $n_g$ and
$n_\alpha$ are relatively small.

In reaching the above equation, we have approximated the
hard-scattering integral $F(p_T)$ that may not be exactly in the form
of $1/[1+m_T/m_{T0}]^n$ into such a form.  It is easy then to see that
upon matching $F(p_T)$ with $A/[1+m_T/m_{T0}]^n$ according to some
matching criteria, the hard-scattering integral $F(p_T)$ will be in
excess of $1/[1+m_T/m_{T0}]^n$ in some region, and will be  in deficit in
some other region.  As a consequence, the ratio of the hard-scattering
integral $F(p_T)$ to the fitting $1/[1+m_T/m_{T0}]^n$ will oscillate
as a function of $p_T$.  This matching between the physical
hard-scattering outcome that contains all physical effects with the
approximation of Eq.\ (\ref{40}) may be one of the origin of the
oscillations of the experimental fit with the non-extensive
distribution (as can be seen below in Fig.\ 3).

\section{ Nonextensive Distribution as a Lowest-Order Approximation of the 
Hard-scattering Integral }

In the hard-scattering integral Eq. (\ref{40}), if we identify
\begin{eqnarray}
n \to \frac{1}{q-1}~~~{\rm and~~}m_{T0} \to \frac{T}{q-1}=nT,
\end{eqnarray}
and consider produced particles to be relativistic so that
$m_T$$\sim$$ E_T$$ \sim $$p_T$ and $E_T$$\sim$$ E$ 
at mid-rapidity, then we will get the nonextensive
distribution of Eq.\ (2) as the lowest-order approximation for the
QCD-based hard-scattering integral.

The convergence of Eq.\ (\ref{40}) and Eq.\ (2) can be considered from
the viewpoint of the reduction of a microscopic description to a
statistical-mechanical description.  From the microscopic perspective,
the hadron production in a $pp$ collision is a very complicated
process, as evidenced by the complexity of the evolution dynamics in
the evaluation of the $p_T$ spectra in explicit Monte Carlo programs,
for example, in \cite{PYTHIA,HERWIG,ARIADNE}.  If one starts from the
initial condition of two colliding nucleons, there are many
intermediate and complicated processes entering into the dynamics,
each of which contain a large set of microscopic and stochastic
degrees of freedom.  Along the way, there are stochastic elements in
the picking of the degree of inelasticity, in picking the colliding
parton momenta from the parent nucleons, the scattering of the
partons, the showering evolution of scattered partons, the
hadronization of the fragmented partons.  Some of these stochastic
elements cannot be definitive and many different models, sometimes with
untestable assumptions, have been put forth.  In spite of all these
complicated stochastic dynamics, the final result of Eq.\ (\ref{40})
of the single-particle distribution can be approximated to depend only
on three degrees of freedom, after all is done, put together, and
integrated.  The simplification can be considered as a ``no hair"
reduction from the microscopic description to nonextensive statistical
mechanics in which all the complexities in the microscopic description
``disappear" and are subsumed behind the stochastic processes and
integrations.  In line with statistical mechanics and in analogy with
the Boltzmann-Gibbs distribution, we can cast the hard-scattering
integral in the non-extensive form in the lowest-order approximation
as \cite{CTWW}\footnote{We are adopting the convention of setting both Boltzmann
  constant~$k_B$ and the speed of light~$c$ to be unity.}
\begin{eqnarray}
&&\hspace*{-1.4cm}\frac{dN}{dy d\boldsymbol{p}_T }\biggr |_{y\sim 0} = \frac{1}{2\pi p_T}
\frac{dN}{dy dp_T} \biggr |_{y\sim 0} = A e_q^{-E/T} ,
\label{qexponential}\\
&&\hspace*{-1.0cm}e_q^{-E/T}  \equiv \left[ 1 - \left(1-q\right) E/T  \right]^{1/(1-q)},~~
e_1^{-E/T} =e^{-E/T},\nonumber
\end{eqnarray}
where $E$=$\sqrt{m^2+{\bb p}^2}$ and $E$=$E_T$=$m_T$ at $y$=0.  Here,
the parameter $q$ is related physically to the power index $n$, the
parameter $T$ related to $m_{T0}$ and the average transverse momentum,
and the parameter $A$ related to the multiplicity (per unity rapidity)
after integration over $p_T$.  Given a physically determined invariant
cross section in the log-log plot of the cross section as a function
of the transverse hadron energy as in Fig. 3, the slope at large $p_T$
gives the power index $n$ (and $q$), the average of $E_T$ gives $T$
(and $m_{T0}$), and the integral over $p_T$ gives $A$.

Fig. 3 gives the comparisons of the results from
Eq.\ (\ref{qexponential}) with the experimental $p_T$ spectra at
central rapidity obtained by different Collaborations
\cite{CMS,ATLAS,ALICE}.  In these calculations, the effective
temperature parameter is set equal to $T$=0.13 GeV, and the parameters
of $A$, $q$ and the corresponding $n$ are given in Table 2.  The
dashed line (an ordinary exponential of~$E_T$ for $q\to$ 1)
illustrates the large discrepancy if the distribution is described by
Boltzmann-Gibbs distribution.  The results in Fig. 3 shows that
Eq.\ (\ref{qexponential}) adequately describes the hadron $p_T$
spectra at central rapidity in high-energy $pp$ collisions.  We verify
that~$q$ increases slightly with the beam energy, but, for the present
energies, remains always $q\simeq 1.1$, corresponding to a power index
$n$ in the range of 6-8 that decreases as a function of $\sqrt{s}$.

\vspace{5mm}
\begin{figure}
\centering
\includegraphics[scale=0.32]{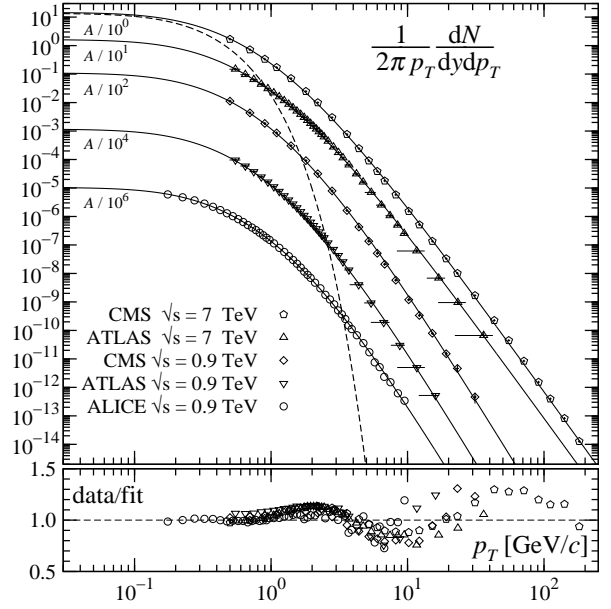}
\caption{Comparison of Eq.\ (\ref{22}) with the experimental
  transverse momentum distribution of hadrons in~$pp$ collisions at
  central rapidity $y$. Herein the temperature is set to be the same
  for all curves and equal $T = 0.13$ GeV, and the normalization
  constant in units of {GeV}$^{-2}/c^3$.  The corresponding
  Boltzmann-Gibbs (purely exponential) fit is illustrated as the
  dashed curve. For a better visualization both the data and the
  analytical curves have been divided by a constant factor as
  indicated. The ratios data/fit are shown at the bottom, where a
  roughly log-periodic behavior is observed on top of the
  $q$-exponential one. Data are taken from
  \cite{CMS,ATLAS,ALICE}.  \label{F4} }
\end{figure}
\begin{table}[h]
\centering \caption{Parameters used to obtain fits presented in
  Fig. \ref{F4} where we have used $T$=0.13 GeV.  The values of $A$ is
  in units of {GeV}$^{-2}/c^3$.  } \label{Tab:LC_fits}
\begin{tabular}{|c|c|c|c|c|}
\hline \!\!Collaboration \!\!& $\sqrt{s}$ [TeV] & $A$ & $q$ & \!\!$n$=1/$(q-1)$\!\!

\\\hline
CMS \cite{CMS}   & $7$   & $38$ & $1.150$  & 6.67\\
ATLAS \cite{ATLAS} & $7$   & $43$ & $1.151$ & 6.62 \\
CMS \cite{CMS}   & $0.9$ & $30$ & $1.127$ & 7.87 \\
ATLAS \cite{ATLAS} & $0.9$ & $32$ & $1.124$ & 8.06 \\
ALICE \cite{ALICE} & $0.9$ & $27$ & $1.124$ & 8.06 \\
 \hline
\end{tabular}
\end{table}

What interestingly emerges from the analysis of the data in
high-energy~$pp$ collisions is that the good agreement of the present
phenomenological fit extends to the whole~$p_T$ region (or at least
for~$p_T$ greater than $0.2\,\textrm{GeV}/c$, where reliable
experimental data are available)~\cite{Won12}. This is being achieved
with a single nonextensive distribution.  On the other hand,
theoretical analysis demonstrates that the hard-scattering integral
can be written as a non-extensive distribution with only three degrees
of freedom, in the lowest-order approximation.  It is reasonable to
infer that the dominant mechanism of hadron production over the whole
range of $p_T$ at central rapidity and high energies is the
hard-scattering process.

The dominance of hard-scattering also for the production of low-$p_T$
hadron in the central rapidity region is supported by two-particle
correlation data where the two-particle correlations in minimum
$p_T$-biased data reveals that a produced hadron is correlated with a
``ridge" of particles along a wide range of~$\Delta\eta$ on the
azimuthally away side centering around $\Delta\phi\sim
\pi$~\cite{STAR,Abe12,Ray11}. The $\Delta\phi\sim \pi$ (back-to-back)
correlation indicates that the correlated pair is related by a
collision, and the $\Delta\eta$ correlation in the shape of a ridge
indicates that the two particles are partons from the two nucleons and
they carry fractions of the longitudinal momenta of their parents,
leading to the ridge of~$\Delta \eta$ at $\Delta\phi\sim \pi$.

\section{Conclusions and Discussions}

Particle production in high-energy $pp$ collisions at central rapidity
is a complex process that can be viewed from two different and
complementary perspectives.  On the one hand, there is the successful
microscopic description involving perturbative QCD and nonperturbative
hadronization at the parton level where one describes the detailed
mechanisms of parton-parton hard scattering, parton structure
function, parton fragmentation, parton showering, the running coupling
constant and other QCD processes.  On the other hand from the
viewpoint of statistical mechanics, the single-particle distribution
can be cast into a form that exhibit all the essential features of the
process with only three degrees of freedom.  The final result of the
process can be summarized, in the lowest-order approximation, by a
power index $n$ which can be represented by a nonextensivity
parameter $q$=$(n+1)/n$, the average transverse momentum $m_{T0}$
which can be represented by an effective temperature $T$=$m_{T0}/n$,
and an multiplicity constant $A$ that is related to the multiplicity
per unit rapidity when integrated over $p_T$.  Such a reduction from
microscopic description to a statistical mechanical description can be
shown both from theoretical considerations by obtaining a simplified
and approximate hard-scattering integral, and also by comparing with
experimental data.  In the process, we uncover the dominance of the
hard-scattering hadron-production and the approximate validity of a
``no-hair" statistical-mechanical description for the whole transverse
momentum region in $pp$ collision at high-energies. We emphasize also
that, {\it in all cases}, the temperature turns out to be one and the
same, namely $T=0.13\,\textrm{GeV}$.

What we may extract from the behavior of the experimental data is that
scenario proposed in \cite{Michael,H} appears to be essentially
correct excepting for the fact that we are not facing thermal
equilibrium but a different type of stationary state, typical of
violation of ergodicity (for a discussion of the kinetic and effective
temperatures see \cite{ET,overdamped})).

As a concluding remark, we note that the data/fit plot in the bottom
part of Fig.~\ref{F4} exhibit an intriguing rough log-periodicity
oscillations, which suggest corrections to the lowest-order
approximation of Eq.\ (\ref{40}) and some hierarchical fine-structure
in the quark-gluon system where hadrons are generated. This behavior
is possibly an indication of some kind of fractality in the
system. Indeed, the concept of \emph{self-similarity}, one of the
landmarks of fractal structures, has been used by Hagedorn in his
definition of fireball, as was previously pointed out in \cite{Beck}
and found in analysis of jets produced in~$pp$ collisions at
LHC~\cite{GWZW}. This small oscillations have already been preliminary
discussed in Section 8 and in~\cite{Wilk1,Wilk2}, where the authors
were able to mathematically accommodate these observed oscillations
essentially allowing the index~$q$ in the very same
Eq.~\eqref{qexponential} to be a complex number\footnote{It should be
  noted here that other alternative to complex~$q$ would be
  log-periodic fluctuating scale parameter~$T$, such possibility was
  discussed in~\cite{Wilk2}.} (see also
Refs.~\cite{logperiodic,Sornette1998}; more details on this
phenomenon, including also discussion of its presence in recent AA
data, can be found in \cite{MR}). \\

Acknowledgments: The research of CYW was supported in part by the
Division of Nuclear Physics, U.S. Department of Energy, and the
research of GW was supported in part by the National Science Center
(NCN) under contract Nr 2013/08/M/ST2/00598 (Polish agency). Two of us
(L.J.L.C. and C.T.) have benefited from partial financial support from
CNPq, Faperj and Capes (Brazilian agencies). One of us (CT)
acknowledges partial financial support from the John Templeton
Foundation.

\end{document}